\documentclass[12pt, epsfig]{article}   
\usepackage{epsfig}     
\usepackage{amssymb}
\usepackage{amsfonts}
\newskip\humongous \humongous=0pt plus 1000pt minus 1000pt

\newif\ifdtup

\catcode`\@=11

\@addtoreset{equation}{section}
\def\theequation{\thesection\arabic{equation}}

\def\@normalsize{\@setsize\normalsize{15pt}\xiipt\@xiipt
\abovedisplayskip 14pt plus3pt minus3pt%
\belowdisplayskip \abovedisplayskip
\abovedisplayshortskip \z@ plus3pt%
\belowdisplayshortskip 7pt plus3.5pt minus0pt}

\def\small{\@setsize\small{13.6pt}\xipt\@xipt
\abovedisplayskip 13pt plus3pt minus3pt%
\belowdisplayskip \abovedisplayskip
\abovedisplayshortskip \z@ plus3pt%
\belowdisplayshortskip 7pt plus3.5pt minus0pt
\def\@listi{\parsep 4.5pt plus 2pt minus 1pt
     \itemsep \parsep
     \topsep 9pt plus 3pt minus 3pt}}

\relax

\catcode`@=12

\catcode`\@=11

\def\section{\@startsection{section}{1}{\z@}{3.5ex plus 1ex minus
   .2ex}{2.3ex plus .2ex}{\large\bf}}

\def\thesection{\arabic{section}.}

\def\appendix{\setcounter{section}{0}
 \def\thesection{Appendix \Alph{section}:}
 \def\theequation{\Alph{section}.\arabic{equation}}}

\def\YGrule{0.4}   
\def\YGbox{6.5}    
\def\SymBoxes#1#2#3#4{\newdimen\un@t \un@t#3%
\raisebox{#1}{\rule{#2\un@t}{#4}\hskip-#2\un@t
\@tempdimb\un@t \advance\@tempdimb by-#4\@tempcntb#2\relax%
\@whilenum{\@tempcntb>0}\do{
\rule{#4}{\un@t}\hskip\@tempdimb \advance\@tempcntb by\m@ne}%
\hskip-#2\un@t \rule[\un@t]{#2\un@t}{#4}%
\rule[\un@t]{#4}{#4}\hskip-#4
\rule{#4}{\un@t}}\hskip-#4}                
\def\Young{\@ifnextchar[{\@Young}{\@Young[0]}}
\def\@Young[#1]#2{\newdimen\YG@unit \YG@unit\YGbox pt%
\newdimen\h@ight \h@ight#1\YG@unit \@tempcnta-1\relax
\@tfor\c@ount:=#2\do{\advance\@tempcnta by\@ne}
\@tempdima\@tempcnta\YG@unit%
\advance\h@ight by\@tempdima\relax     
\@tfor\c@ount:=#2\do{\SymBoxes{\h@ight}{\c@ount}{\YG@unit}{\YGrule pt}%
\@tempdima-\c@ount\YG@unit \hskip\@tempdima%
\advance \h@ight by -\YG@unit}         
\@tempdima\YG@unit \multiply\@tempdima by\@car#2\@nil %
\hskip\@tempdima}                      
\def\YoungTab{\@ifnextchar[{\@YoungIdx}{\@YoungIdx[0]}}
\def\@YoungIdx[#1]{\@ifnextchar[{\@iYoungIdx[#1]}{\@iYoungIdx[#1][\@empty]}}
\def\@iYoungIdx[#1][#2]#3{%
\newdimen\YG@unit \YG@unit\YGbox pt\newdimen\YG@rule \YG@rule \YGrule pt

\newcount\c@ount \c@ount\z@ \newdimen\skip@wd \unitlength\@ne pt
\newdimen\h@ight \h@ight#1\YG@unit \@tempcnta\m@ne\relax
\@tfor\d@um:=#3\do{\advance\@tempcnta by\@ne}
\@tempdima\@tempcnta\YG@unit%
\advance\h@ight by\@tempdima\relax
\@tfor\@idxlist:=#3\do{
\@tempcnta\z@\hskip.5\YG@rule\relax
\@for\@idx:=\@idxlist\do{
\raisebox{\h@ight}{\makebox(\YGbox,\YGbox){#2$\@idx$}}
\advance\@tempcnta by\@ne}\hskip-.5\YG@rule%
\@tempdima-\@tempcnta\YG@unit \hskip\@tempdima%
\ifnum\c@ount=\z@ \skip@wd-\@tempdima\fi \relax
\SymBoxes{\h@ight}{\@tempcnta}{\YG@unit}{\YG@rule}%
\hskip\@tempdima \advance\h@ight by -\YG@unit
\advance\c@ount by\@ne}
\hskip\skip@wd}                      

\begin{document}

\newcommand{\beq}{\begin{equation}}
\newcommand{\eeq}{\end{equation}}
\newcommand{\bea}{\begin{eqnarray}}
\newcommand{\eea}{\end{eqnarray}}
\newcommand{\beas}{\begin{eqnarray*}}
\newcommand{\eeas}{\end{eqnarray*}}
\newcommand{\defi}{\stackrel{\rm def}{=}}

\newcommand{\non}{\nonumber}
\newcommand{\bquo}{\begin{quote}}
\newcommand{\enqu}{\end{quote}}
\def\de{\partial}
\def\Tr{ \hbox{\rm Tr}}
\def\const{\hbox {\rm const.}}
\def\o{\over}
\def\im{\hbox{\rm Im}}  
\def\re{\hbox{\rm Re}}
\def\bra{\langle}\def\ket{\rangle}
\def\Arg{\hbox {\rm Arg}}
\def\Re{\hbox {\rm Re}}
\def\Im{\hbox {\rm Im}}
\def\diag{\hbox{\rm diag}}
\def\longvert{{\rule[-2mm]{0.1mm}{7mm}}\,}

\begin{titlepage}
{\hfill     IFUP-TH 2002/46}
\bigskip
\bigskip
\bigskip
\bigskip

\begin{center}
{\Large  {\bf
  ALMOST CONFORMAL VACUA AND CONFINEMENT    
} }
\end{center}
\vspace{1em}
\begin{center}
{\large \bf Roberto AUZZI $^{(1,3)}$ , Roberto GRENA $^{(2,3)}$   \\ and    \\   \vskip 0.15cm   Kenichi  KONISHI $^{(2,3)}$ }
\end{center}
\vspace{1em}
\begin{center}
{\it
Scuola Normale Superiore - Pisa $^{(1)}$   \\ 
 Piazza dei Cavalieri 7,   Pisa, Italy  \\
Dipartimento di Fisica   ``E. Fermi"  -- Universit\`a di Pisa $^{(2)}$ \\
Via Buonarroti, 2,   Ed. C, 56127  Pisa, Italy \\
Istituto Nazionale di Fisica Nucleare -- Sezione di Pisa $^{(3)}$ 
\\
     Via Buonarroti, 2,   Ed. C, 56127  Pisa, Italy}

{ \bf  konishi@df.unipi.it, grena@df.unipi.it, auzzi@sns.it}
\end {center}

\vspace{3em}
\noindent
{\bf Abstract:}

 {  Dynamics of confining vacua which appear as  deformed superconformal theory
with a non-Abelian gauge symmetry,     is studied   by taking a concrete example of the sextet vacua 
of ${\cal N}=2$, $SU(3)$  gauge theory with $n_f=4$, with  equal  quark  masses.
  We show that  the low-energy ``matter"  degrees of freedom of this theory consist of  four  magnetic monopole doublets
of the low-energy effective  $SU(2)$ gauge group, one dyon doublet, and one electric doublet.  We find  a mechanism of 
cancellation of the beta function, which naturally but nontrivially generalizes that of Argyres-Douglas.  Study of our    SCFT
theory  as a limit of six  colliding ${\cal N}=1$
  vacua,     suggests  that the confinement  in the present theory occurs in an  essentially different manner
from those  vacua with dynamical Abelianization,  
and involves strongly interacting non-Abelian magnetic monopoles.  
}

\vfill

\begin{flushright}
November  2002
\end{flushright}
\end{titlepage}

\bigskip

\section{Introduction}

The true mechanism of confinement in QCD is still covered in a  mystery,
 in spite of considerable amount of work dedicated   to this problem.
In a recent series of papers \cite{CKMP}  the mechanism of confinement and of
dynamical symmetry breaking has been
studied  in some detail    in the  context of ${\cal N}=2$  supersymmetric quantum chromodynamics (SQCD),   
broken softly to  ${\cal N}=1$.    An  interesting fact emerged from these analyses:
non-Abelian monopoles of the type studied by
Goddard-Olive-Nuyts \cite{GNO}  make appearance as low-energy degrees of freedom, and play a central
role in the infrared dynamics in some of the
confining vacua \cite{BK}.

The most intriguing type of vacua among those found in \cite{CKMP}, however,   are   the ones
based on  deformation (perturbation) of  superconformal field
theories (SCFT) \cite{SCF,Eguchi}.       The low-energy degrees of freedom involve relatively nonlocal
dyons and there is no local effective Lagrangian describing
them.  Upon perturbation - small adjoint mass - confinement and dynamical
symmetry breaking ensue,  as can be  demonstrated indirectly
through various  considerations,  such as  the study  of the large $\mu$ effective action,
supersymmetry and holomorphy,   and the vacuum counting.

Though far-fetched it  might sound,
confinement {\it is}    described precisely this way in many systems within the
context of ${\cal N}=2$   supersymmetric gauge theories
with $n_f$ hypermultiplets.   Examples are
the $r= {n_f \o 2}$ vacua \footnote{We recall that confining ${\cal N}=1$ vacua arising by perturbing the ${\cal N}=2$  
$SU(n_c)$   gauge theories with the adjoint mass term are classified by different  effective gauge groups, 
$SU(r) \times U(1)^{n_c-r-1}. $    }  in $SU(n_c)$  theories with equal mass flavors,  and {\it all }  
of the confining vacua of
$USp(2n_c)$ and $SO(n_c)$  theories with matter fields with zero bare masses \cite{CKMP}.   
In contrast,
the Abelian dual  superconductor mechanism \cite{TH,MN}   is realized  rather as exceptional cases:    $r=0$ or $r=1$  vacua in $SU(n_c)$  
theory,  or in pure ${\cal N}=2$ Yang-Mills theories, i.e., $n_f=0$.

 Whether a similar mechanism is at work in the standard Quantum Chromodynamics (QCD)  is not known.      The phenomenon
is deep, though,  and   in our opinion  deserves  
a closer look than has been given  so far.

In this paper,   we make a first step  in that direction.  We study the nature of
the  low-energy degrees of freedom in vacua in which
confinement appears to be caused by a collaboration of relatively non-local
light monopoles and dyons.   As an example,  we study the $r=2$ vacua of 
${\cal N}=2$   supersymmetric  $SU(3)$   gauge theory 
with $n_f=4 $ hypermultiplets.  
We show that  the low-energy degrees of freedom of this theory consist of four monopole doublets of the effective 
$SU(2)$  gauge group,  one dyon doublet, and one electric doublet.     We show   how they conspire to 
give a vanishing beta function,  generalizing the Argyres-Douglas mechanism \cite{AD,BiFe}    in a nontrivial manner. 

Study of the superconformal theory as a limit of a number of  colliding  ${\cal N}=1$  vacua, as realized by first considering the theory at unequal
quark masses and then  taking the limit of equal masses,  indicates   that  confinement mechanism at the almost conformal vacua such as the one under
consideration, is essentially distinct from the one at work  in those vacua where confinement is due to the  condensation of weakly coupled magnetic 
particles as first found in Ref. \cite {SW1,SW2}.

\section {The Sextet   ($r=2$)    Vacua of  $SU(3),$  $n_f=4$  Gauge Theory }

We study  the sextet vacuum ($r=2$ vacua)  of    $SU(3)$  gauge theory with four flavors
of quarks.   The
Seiberg-Witten curve \cite{SW1,SW2}   of this theory is  (by setting $2 \Lambda=1$) \cite{curves}  
\beq  y^2 =  \prod_{i=1}^{3}  (x-\phi_i)^2 -  \prod_{ a=1}^4    (x+m_a ) \equiv    (x^3  -   U  x  -
V )^2  -  \prod_{ a=1}^4    (x+m_a ).
\label{SWbis} \eeq
For {\it equal bare quark masses}    ($m_a=m$),   it simplifies: 
\beq  y^2 =  \prod_{i=1}^{3}  (x-\phi_i)^2 -    (x+m )^4    \equiv    (x^3  -   U  x  -
V )^2  -  (x+m )^4.  
\label{SW} \eeq
The sextet vacua correspond to  the point,  $diag\,  \phi =  (-m, -m,  2m )$,
\i.e.,
\beq    U=   { 3 m^2}; \qquad   V= 2 m^3,
\label{Sing}\eeq
 where the curve exhibits  a singular behavior,
\beq   y^2 \propto   (x+m)^4
\eeq
corresponding to the unbroken $SU(2)$  symmetry.
This singularity  splits to six separate singularities when the quark masses are
taken to be slightly unequal and generic.

\begin{figure}[h]
\begin{center}
\epsfig{file=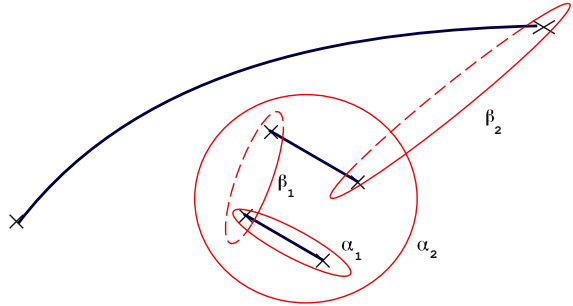,  width=8cm}
\end{center}    
\caption{ \footnotesize  The canonical homology cycles  together with the cuts (thick lines), in the double-sheeted Riemann surface
representation of the curve (\ref{SWbis}).    
}\label{branchp}  
\end{figure}

At a generic point $(U,V)$ near  (\ref{Sing})  the right hand side of the curve
(\ref{SW})  has  six square-root  branch points, four of which are near
$x=-m$, the other two are at separate points of ${\cal O}(1)$.  We  draw the canonical cycles as
in Fig.\ref{branchp}
and define
\beq    a_{D1}  =  \oint_{\alpha_1}  \lambda, \qquad  a_{D2}  =
\oint_{\alpha_2}  \lambda, \qquad
 a_{1}  =  \oint_{\beta_1}  \lambda, \qquad  a_{2}  =  \oint_{\beta_2}  \lambda,
\eeq where the (meromorphic)  one-form $\lambda$ is given by \cite{curves}
\beq
\lambda =  { x \o 2 \pi}  d\, \log{  \prod (x -\phi_i) -y \o  \prod (x -\phi_i)+y  }. 
 \eeq 
The masses  of the particles having the magnetic and
electric quantum numbers  $(g_{1}, g_{2}; q_{1}, q_{2})$  are then
given by the formula \beq  M_{ (g_{1}, g_{2};  q_{1}, q_{2})} =
\sqrt 2 \,  | g_{1}\, a_{D1} +  g_{2}\,a_{D2}  +    q_{1}\,a_{1} +
q_{2} \, a_2 |.    \eeq

\subsection {Singularity Structure near the Conformal Point and Low-energy
Degrees of Freedom}

The conformally invariant vacua   occur  at the points where more than one
singularity loci in the $u-v$ plane,
corresponding to relatively nonlocal massless states,  coalesce \cite{SCF,Eguchi}.
The nature  of  the  low-energy degrees of freedom at the SCFT vacua can be determined by
studying the monodromy matrices around each
of the singularity curves.  In the case of the $r=2$ vacua of $SU(3)$ theory
with $n_f=4$,  it is necessary to study the
behavior of the theory near the point Eq.(\ref{Sing}).  Set
\beq   U=    { 3 m^2} +u, \qquad V=  2 m^3 + v,   \qquad  x +m \to   x.
\eeq
The discriminant of the curve factorizes \cite{Eguchi}  as
\beq    \Delta=  \Delta_s \, \Delta_{+} \,  \Delta_{-},
\eeq
where the squark singularity\footnote{At large $m$ hence at large  $U$ and $V$
it represents massless quarks and squarks;   as is well known,
at small  $m$ it  becomes  monopole singularity, due to the fact that the corresponding singularity  goes under certain cuts produced
by other singularities \cite{BF,Cp}. }    corresponds to the factor
\beq  \Delta_s =  (m \, u - v)^4;
\eeq
where the fourth zero represents  the flavor   multiplicity $n_f=4$;
\bea  \Delta_{+} &=&  4\,m\,u + 36\,m^2\,u + 108\,m^3\,u + 108\,m^4\,u + u^2 +
24\,m\,u^2 + 36\,m^2\,u^2  +   \non \\
 &&   4\,u^3 - 4\,v - 36\,m\,v - 108\,m^2\,v -
  108\,m^3\,v - 18\,u\,v - 27\,v^2,
\eea
\bea  \Delta_{-}&=&  -4\,m\,u + 36\,m^2\,u - 108\,m^3\,u + 108\,m^4\,u + u^2 -
24\,m\,u^2 + 36\,m^2\,u^2  +  \non \\
&& 4\,u^3 + 4\,v - 36\,m\,v + 108\,m^2\,v -
  108\,m^3\,v + 18\,u\,v - 27\,v^2,
\eea represent the loci where  some other  dyons become
massless. The    equations $\Delta_{\pm}=0  $  can be  approximated by
\beq v = m u + \frac{1}{12 m + 4} u^2+O(u^3),\eeq
 \beq v = mu + \frac{1}{12 m-4} u^2+O(u^3) \eeq
at small $m$.  
Thus at sufficiently small $u,v,m$,  the three equations
$\Delta_s=0, \Delta_{+}=0, \Delta_{-}=0$  can be replaced by \beq
v=m\,u, \qquad  v=  m\,u +    \frac{u^2}{4}, \qquad v=  m\,u -
\frac{u^2}{4}. \eeq These three (complex) curves  meet at $u=v=0$
tangentially.    For the purpose of studying the topological
feature of the three curves one can further  rescale $u,v$ by   $u =
 m \, {\tilde u}$,  $v=   m^2 \, {\tilde v }$  so that  the three curves are now
\beq   {\tilde v }={\tilde u}, \qquad  {\tilde v }=  {\tilde u}+    \frac{{\tilde u}^2}{4}, \qquad {\tilde v }= {\tilde u} -
\frac{{\tilde u}^2}{4}. \label{curves}\eeq

\begin{figure}[h]
\begin{center}
\epsfig{file=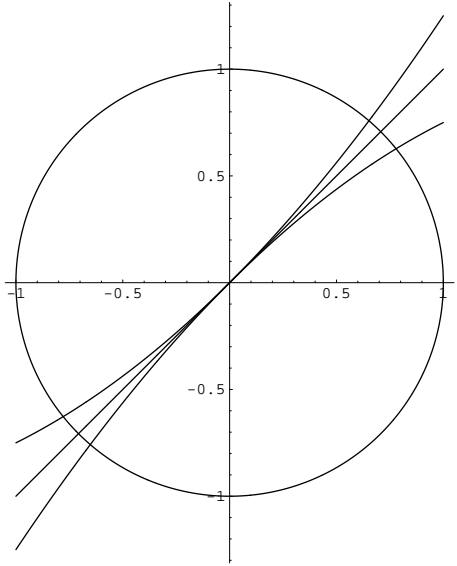,  width=6cm}
\end{center}
\caption{\footnotesize  A two-dimensional  ($\Re \, {\tilde u} - \Re \,{\tilde v}$)   projection of the  
curves (\ref{curves}) and  the sphere  $S^3,$ (\ref{sphere}).    }\label{twodim}
\end{figure}

In order to define uniquely  the monodromies around the three curves near the
SCFT  point  we consider
the intersections of these curves
with the $S^3$   sphere
\beq    |{\tilde u}|^2 + |{\tilde v}|^2 =1. \label{sphere}
\eeq
   A two-dimensional ($\Re \, {\tilde u} - \Re \,{\tilde v}$)  projection of the curves
(\ref{curves}), (\ref{sphere}),
is shown  in Fig.\ref{twodim}.

\begin{figure}[h]
\begin{center}
\epsfig{file=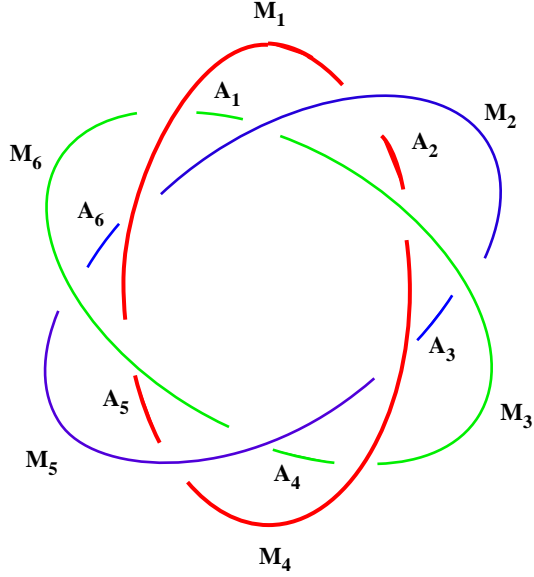,    width=7 cm}
\end{center}
\caption{\footnotesize   Zero loci of the discriminant of the curve of 
${\cal N}=2,$ $\,SU(3)$, $\,n_f=4$ theory at small $m$.}
\label{knot}
\end{figure}

These intersections form    one-dimensional closed curves  (on the two
dimensional projection in Fig.\ref{twodim} just two points
are visible for each such intersection curve).
We consider  various closed curves lying on $S^3$    and  encircling the curves
(\ref{curves}) at various points.
It is convenient to make first  a stereographic projection  from  $S^3  \to
R^3$    by
\beq    X=  {w_1 \o  1 - w_4}; \quad  Y=  {w_2 \o  1 - w_4}; \quad Z=  {w_3 \o
1 - w_4},
\eeq where \beq     {\tilde u}= w_1 + i  \, w_2; \quad {\tilde v}= w_3 + i  \, w_4,
\eeq
after which   the intersection curves  take the form of the three
linked rings.
   It can be shown that each pair of the rings are linked non-trivially   with
linking number two \footnote{The linking number is defined
by
$$   N=  { 1\o 4\pi}  \oint \oint dx_{i} \, dy_j \,  \epsilon_{ijk}  {  (x-
y)_{k} \o    ((x-y)^2)^{3/2}}
$$
where the integrations are along  the two rings.
}.  Furthermore,  the linking   between  all three pairs of the rings can be shown to occur    in  the
same direction.   Topologically, the three intersection curves look
like those in Fig. \ref{knot}.

 \begin{figure}[h]
\begin{center}
\epsfig{file=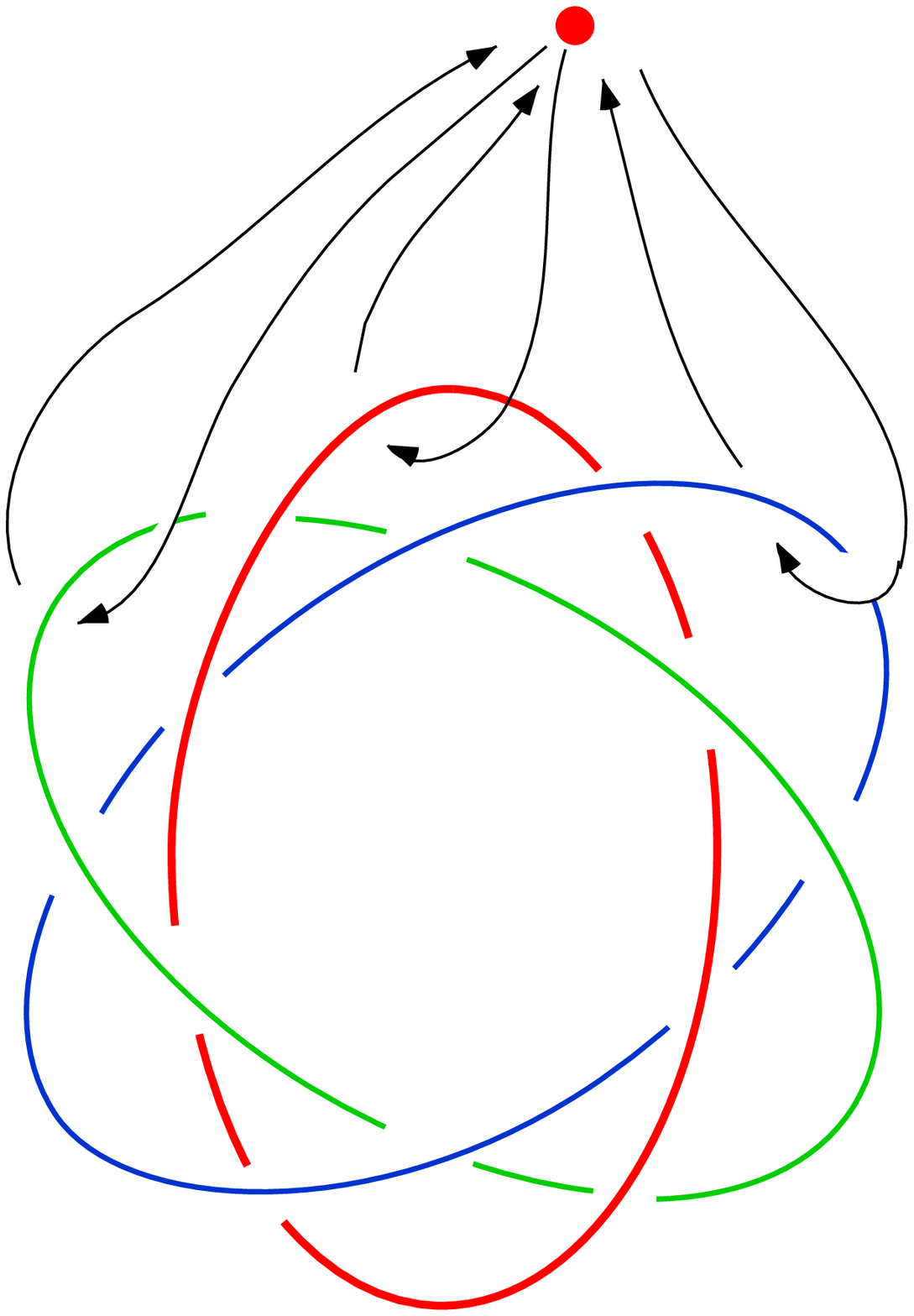,   width=6cm}
\end{center}
\caption{\footnotesize   A few closed paths around some parts of the linked rings.  }
\label{paths}
\end{figure}

We consider now  various closed paths  in the space $(u,v)$,   starting from  a fixed  reference point
(for instance lying above the page),
encircling various parts of the rings   and coming back to the original  point,
as in the three  paths shown in Fig. \ref{paths}.     These induce the movements of
 the four branch points near the origin  (the slight  movements of the furthest
ones near   $\pm 1$  are irrelevant).
  As the branch points move,   the  integration contours   $\alpha_i$'s and   $\beta_i$'s    get entangled in a non-trivial way.
For example,   in the case of the central closed circuit
of Fig. \ref{paths}  in which the curve ${\tilde v}={\tilde u}$     is encircled once near  ${\tilde u}=1,\,{\tilde v}=1$,
the two branch points closest to the origin  (call $x_1, x_2$)
rotate with respect to each other,   $\Arg (x_1- x_2) $  going through a change
of $4 \pi$.
 The canonical  cycles  go
through the change
\beq   \alpha_1 \to \alpha_1, \qquad  \beta_1 \to   \beta_1 - 4 \alpha_1, \qquad
\alpha_2 \to \alpha_2, \qquad  \beta_2 \to \beta_2.
\eeq
The monodromy transformation is thus
\beq    \pmatrix{a_{D1}  \cr   a_{D2}  \cr  a_1 \cr  a_2 }   \to  M_1
\pmatrix{a_{D1}  \cr   a_{D2}  \cr  a_1 \cr  a_2 },
\qquad
  M_1= {\tilde M}_1^4, \quad  \, {\tilde M}_1=\pmatrix   { 1 & 0& 0  & 0 \cr
0& 1 & 0& 0 \cr   -1 & 0& 1& 0\cr
 0& 0& 0&  1  }   \label{M1}  \eeq
From the   well-known formula \cite{curves}
\beq   M=  \pmatrix  {{\bf 1} +  {\vec q} \otimes  {\vec g} &   {\vec q} \otimes
{\vec q}  \cr
  - {\vec g} \otimes  {\vec g}  &     {\bf 1} -    {\vec g} \otimes  {\vec q} }
\label{charges}    \eeq
one concludes that the (four)  massless particles   at the singularity ${\tilde v}={\tilde u}$
have charges
\beq     (g_{1}, g_{2}; q_{1}, q_{2})  =   (1,0;0,0).   \eeq
Analogously, the monodromy transformations around  the  ${\tilde v }=  {\tilde u}+    \frac{{\tilde u}^2}{4}, \quad {\tilde v }= {\tilde u} -
\frac{{\tilde u}^2}{4}$  are determined to be
\beq
M_{2}  =  \pmatrix   { -1 & 0& 1  & 0 \cr    0& 1 & 0& 0 \cr   -4 & 0& 3& 0\cr
 0& 0& 0&  1  }, \qquad  M_{6}  =  \pmatrix   {1 & 1& 1  & 0 \cr    0& 1 & 0& 0
\cr   0 & 0& 1& 0\cr
 0& -1& -1 &  1  },  \eeq
respectively.

In principle, this procedure could be carried out for all other parts of the
rings, but there is a more systematic procedure, which takes the
symmetry of the system into account. Indeed,  various monodromy transformations
are related by the conjugations,
\bea   && M_1=  M_6^{-1}   A_5  M_6, \quad     A_2 =  M_2^{-1}  M_1  M_2, \quad
M_4 =  M_3^{-1}   A_2  M_3, \quad   A_5 =M_5^{-1}   M_4  M_5,
\non\\  &&  M_2 = M_1^{-1}   A_6  M_1, \quad     A_3 =  M_3^{-1}  M_2  M_3,
\quad   M_5 =  M_4^{-1}   A_3  M_4, \quad   A_6=  M_6^{-1}   M_5  M_6,
\non \\  &&  M_3 = M_2^{-1}   A_1  M_2, \quad     A_4 =  M_4^{-1}  M_3  M_4,
\quad   M_6 =  M_5^{-1}   A_4  M_5, \quad   A_1 =  M_1^{-1}   M_6  M_1
\non 
 \\ \label{conju}  \eea  
as can be easily verified  by   looking at  Fig. \ref{knot}.
By knowing any three of them, for instance $M_1, M_2, M_6$  above,   these
relations yield uniquely all
 the other monodromy matrices.  The twelve
monodromy matrices $M_1 \sim M_6,$
$A_1 \sim A_6 $   determined this way    are listed in Appendix A.
The formula (\ref{charges})   then gives the charges
\bea  && M_1: \, (1,0;0,0)^4, \quad M_4:\,\, (-1,1; 0,0)^4, \quad  M_2: \, (-2,0;1,0), \quad M_5:\,\, (2,-2; -1,0), 
\non \\  &&  A_2:\,\,  (-1, 0; 1,0)^4, \quad  A_5:\,\, (1,-1; -1,0)^4,   \quad A_3:\,\, (-
2, 2; -1,0), \quad  A_6:\,\, (2,0; 1,0),
\non \\ && M_3: \, (0,1;-1,0), \quad M_6:\,\, (0,1; 1,0), \quad A_4:\,\, (4, -3;
-1,0), \quad  A_1:\,\, (-4,1; 1,0),
\non  \\ 
\label{chargesbis} \eea where  the superscript $4$  for $M_1, \, M_4,\, A_2, \, A_5$ 
indicates the fact that these charges appear four times (the
monodromy matrix  being the  fourth power of an elementary
monodromy matrix, corresponding to these charges).   Note that all
of them have a vanishing charge    $q_2$;  thus  in the limit $u
\to 0, \,\, v\to 0$  (where $\alpha_1, \beta_1, \,\alpha_2$ cycles
shrink to zero cycles)  all of them would become massless
simultaneously.

\begin{figure}[h]
\begin{center}
\epsfig{file=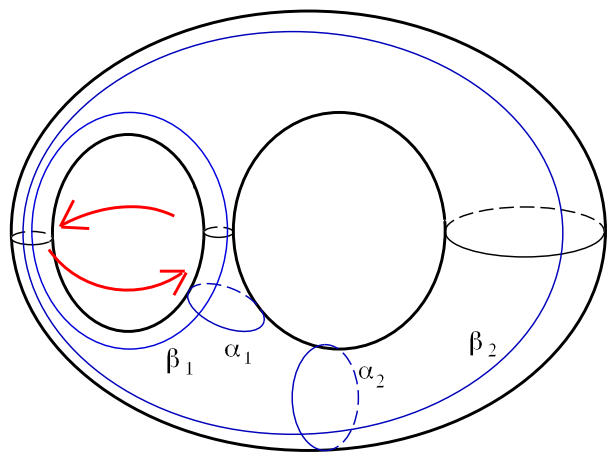,    width=8cm}  
\end{center}
\caption{\footnotesize   Exchanging the two equivalent necks of the bi-torus   }
\label{necks}
\end{figure}

 We wish to know how these particles  behave under   the  $SU(2) \times U(1) $
group  which is unbroken  at $u=v=0$.  Although
we know that the charges (\ref{chargesbis})  refer to a (magnetic or  eletric)
$U(1)^2$  subgroup of $SU(3)$,  it is not
a priori clear how it  is related the unbroken $SU(2) \times U(1) $
group.
A useful element is the transformation property under the canonical base change,
\beq   \alpha_1 \to \alpha_2 - \alpha_1;  \qquad \beta_1 \to - \beta_1; \qquad
 \alpha_2 \to  \alpha_2;\qquad  \beta_2 \to \beta_1 + \beta_2.
\eeq
Note that the intersection numbers $ \alpha_i \# \beta_j = \delta_{ij}$ are
maintained.  Geometrically, it corresponds to the
exchange of the two equivalent  ``necks"  of the bi-torus  (Fig.\ref{necks}), and can be
interpreted as an $SU(3)$  transformation in
which   the first and the second component of the fundamental multiplet are
exchanged.

This transformation induces the change of charges \beq    g_1 \to
-  g_1; \qquad g_2 \to g_1 +   g_2; \qquad q_1 \to  -  q_1 +  q_2;
\qquad  q_2 \to q_2. \label{chargech} \eeq An inspection shows
that the charges  (\ref{chargesbis})     are actually paired as
doublets  transforming  into each other under it: \bea  && \,
(1,0;0,0) \leftrightarrow (-1,1; 0,0), \quad \,\, (-1, 0; 1,0)
\leftrightarrow  (1,-1; -1,0), \non \\  &&  (-2,0;1,0)
\leftrightarrow (2,-2; -1,0), \quad  (-2, 2; -1,0) \leftrightarrow
(2,0; 1,0), \non \\ &&  (0,1;-1,0) \leftrightarrow (0,1; 1,0),
\quad  (4, -3; -1,0) \leftrightarrow (-4,1; 1,0).
\label{chargepairs} \eea These pairs of charges can therefore  be
interpreted as belonging to various {\it doublets}  of the
unbroken $SU(2)$ gauge group. It is easy to introduce, accordingly,    the  magnetic
and electric $U_1(1), \,U_2(1) $  charges  such that 
the first is a subgroup of the $SU(2)$ while the second is orthogonal to it:
 \beq Q_1= \pmatrix{1 & 0 & 0 \cr   0 & -1
& 0 \cr 0 & 0 & 0 }, \qquad  Q_2= \pmatrix{\frac{1}{2} & 0 & 0 \cr
0 & \frac{1}{2} & 0 \cr 0 & 0 & -1}.\eeq
 The corresponding
magnetic and electric $U_1(1)$ charges  are \beq   {\tilde  m}_1 =
m_1; \quad {\tilde q}_1 =  q_1 - \frac{1}{2} \, q_2 ; \label{newbas}\eeq 
and 
\beq
{\tilde m}_2 = m_1 + 2 \, m_2; \quad  {\tilde q}_2 =  \frac{1}{2} \, q_2
\label{newbasis}   \eeq respectively. The normalization is chosen such  that the
transformation has determinant
unity   in the   
$\{m_1, m_2;    q_1,   q_2\}$ space.   The charges of the doublets in the new basis are:

\begin{table}[h]     
\begin{center}    
\begin{tabular}  {|l|l|} \hline Matrix & Charge \\
\hline
$M_1,M_4$ & $(\pm 1,1,0,0)^4$  \\ $A_2, A_5$ &$(\pm 1,-1,\mp 1,0)^4$ \\
$M_2,M_5$ &$(\pm 2,2,\mp 1,0)$ \\ $A_3,A_6$ & $(\pm 2,-2,\pm 1,0)$  \\
$M_3,M_6$&$(0,2,\pm 1,0)$  \\ $A_1,A_4$ & $(\pm 4,-2,\mp 1,0)$  \\
\hline
\end{tabular}   
\caption{\footnotesize The charges of the massless doublets in the new basis (\ref{newbas}), (\ref{newbasis}).}     
\label{monopcharges}  
\end{center}
\end{table}
Note that only the members of the same doublet are relatively local, i.e.,  have  a vanishing  relative Dirac unit \cite{TH}   
 \beq  {\cal N}_{D}= \sum_{i=1}^2   (\,g_{A i } \, q_{B i} -  q_{A i} \,  g_{Bi} \,).      
\eeq

The  problem now is to find out  which of these massless   particles are actually present in the low-energy superconformal theory 
defined at  $u=v=0$, and see how the  relatively non-local matter  fields cooperate  to give, together with the (dual) gauge fields and their
superpartners, a  vanishing beta function.

\subsection{Low-Energy Coupling Constant }

The low-energy effective coupling constants  at the superconformal point can be determined by studying the behavior of the curve near that
point. 
For 
$m=0$  we have the following Riemann surface:
 \bea y^2=(x^3-u x -v)^2-x^4=(x^3+x^2-u x -v)(x^3-x^2-u x
-v).\eea 
At   $u=0,v=0$ the  branch points are at: 
\beq    x_1=x_2=x_3=x_4=0,\quad x_5=-1,\quad x_6=1.\eeq The periods
$a_{D1},a_{D2},a_1$ become small: this corresponds to the
non-local charges of the previous section which become massless in
the sextet vacuum.  

A similar degeneration of  genus two Riemann surfaces was studied
in \cite{lebowitz}. The main result derived there  is that in the limit
in which three of the six branch points of a genus two curve
coalesce, the period matrix $\tau_{ij}$ splits as: \bea
\tau_{ij}=\pmatrix { \tau_{11} & 0 \cr
  0&  \tau_{22}  }.\eea
$\tau_{22}$ is the modulus of the ``large"    torus.   If the three  
colliding points coalesce in $a$, and the other three branch
points are respectively at $0,1,\infty$, then \footnote{$\theta_{00}$ and $ \theta_{10}$  are defined   \cite{lebowitz}  as:    
   $\theta_{00}(0,\tau)=\sum_{n=- \infty}^{\infty} e^{i \pi t
n^2 }$ and $\theta_{10}(0,\tau)=\sum_{n=- \infty}^{\infty} e^{i
\pi t (n+\frac{1}{2})^2 }$.    They coincide with   the standard Jacobi Theta functions \cite{WW}:    
  $\theta_{00}(0,\tau) = {\vartheta}_3(0,  | \,  \tau_{11} ); $     $\theta_{10}(0,\tau) = {\vartheta}_2(0,  |  \,\tau). $ }: \bea
a=\frac{\theta_{00}^4(0,\tau_{22})}{\theta_{10}^4(0,\tau_{22})}.\eea
  The modulus of the ``small"  torus $\tau_{11 }$  is well
defined only if in our limit the angles formed by the three 
colliding points are kept constant. If    their complex
coordinates are given by $b,c,d$, one has: 
\bea   
\frac{c-b}{d-
b}=\frac{\theta_{00}^4(0,\tau_{11})}{\theta_{10}^4(0,\tau_{11})} =  { {\vartheta}_3^4(0  \, |  \tau_{11} )
 \o    { \vartheta}_2^4(0\, | \tau_{11}   ) }. \label{relation}\eea
The right hand side is the inverse of the modular $\lambda$ function
$\lambda(\tau) \equiv    {  { \vartheta}_2^4(0\, | \tau   ) 
 \o       { \vartheta}_3^4(0  \, |  \tau   ) }.  $   The relation (\ref{relation}) coincides with the  inversion formula given in \cite{WW}.
See Appendix B.  

\begin{figure}[h]
\begin{center}
\epsfig{file=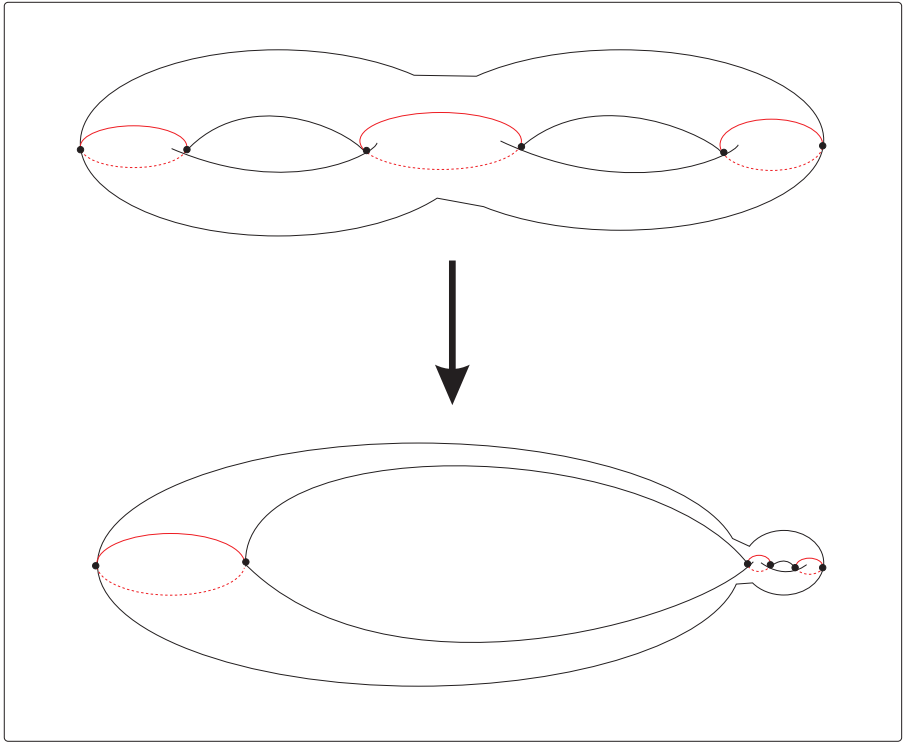,  width=12cm}        
\end{center}  
\caption{\footnotesize The large torus degenerates ($\tau_{22} \to 0$).  The Abelian factor is a trivial infrared free theory.   }
\label{degen}
\end{figure}

In the sextet vacuum we have a special situation: the three
colliding points coalesce with  one of the other three branch
points. So the ``large" torus degenerates ($\tau_{22}=0$), corresponding to the fact that the $U(1)$
factor in the unbroken $SU(2) \times U(1)$ gauge group  is a trivial IR free theory, as at   the  singularities of the  $SU(2)$ Seiberg-Witten 
theory \cite{SW1}.   See Fig. \ref{degen}.

On the other hand,  the shape of the small torus 
$\tau_{11}$ is a function of the relative  angles formed by the four colliding branch points. Let 
$(x_1,x_2,x_3,x_4)$ be  the coordinates of the four colliding points.
By translating them as  $(0,x_2-x_1,x_3-x_1,x_4-x_1)$ and making the 
transformation $x\rightarrow \frac{1}{x}$: we obtain the positions of the three finite branch points (as appropriate for 
using the inversion formula  (\ref{relation}))   at   $(\frac{1}{x_2-x_1},\frac{1}{x_3-x_1},\frac{1}{x_4-x_1})$  and one at infinity.

We must  now identify the three finite branch points  in  (\ref{relation}) with
$\frac{1}{x_2-x_1}$,  $\frac{1}{x_3-x_1}$, $\frac{1}{x_4-x_1}$,   
to determine  $\tau_{11 }$. Notice that the results does not depend
on  the choice  which of the four points is moved to  infinity, but do depend on the way $(b,c,d)$ are identified with
$(\frac{1}{x_2-x_1},\frac{1}{x_3-x_1},\frac{1}{x_4-x_1}).$  This must be done  consistently with the charge assignment, i.e., in
accordance with the choice of the  homology cycles  encircling the branch points,   defining $a_{D i}, a_{i}$   (see Fig.\ref{branchp}).

The main  problem however  is  that of  the uniqueness of the superconformal theory in the limit  $u, v \to 0. $
The behavior  of the branch points (see Eq.(\ref{brnchpts}), Eq.(\ref{branchpoints}) below)   as $u, v \to 0$ clearly shows that 
the ratios  among $(\frac{1}{x_2-x_1},\frac{1}{x_3-x_1},\frac{1}{x_4-x_1})$     depend   on the 
way the limit is approached.   How can one avoid the  arbitrariness of the  value of $\tau_{11}$ in the limit, and 
hence of the superconformal theory defined in such a limit?

In the limit $u, v \rightarrow 0$ the positions of the four
colliding branch points can be determined approximately  from  the following   equation: \bea
(x^2-u x -v)(x^2+u x +v)=0,\eea with the solution: \bea
&&   x_1= \frac{1}{2} (-u+\sqrt{u^2-4 v}); \, \, \, x_2= \frac{1}{2} (-u-\sqrt{u^2-4
v}); \non \\   &&    x_3= \frac{1}{2} (u+\sqrt{u^2+4 v})  ;
\,\,\, x_4= \frac{1}{2} (u-\sqrt{u^2+4 v}). \label{brnchpts} \eea
 Upon introduction of  the variables $\epsilon, \rho,z,w$  following  \cite{AD}, 
 given by: \bea v=\epsilon^2; \qquad    u=\epsilon \rho;
\qquad     x=\epsilon z; \qquad    y=\epsilon^2 w, \eea 
these 
equations become: \beq (z^2-\rho z -1)(z^2+\rho z +1)=0; \label{simm} \eeq
 \bea
&&  z_1=\frac{1}{2} (-\rho+\sqrt{\rho^2-4 }) ; \, \, \,
z_2=\frac{1}{2} (-\rho-\sqrt{\rho^2-4 }); \non \\ 
&&   z_3=\frac{1}{2}  (\rho+\sqrt{\rho^2+4 }); \,\,\,
z_4=\frac{1}{2} (\rho-\sqrt{\rho^2+4 }).  \label{branchpoints} \eea
The  positions of the four branch points depend  thus   only on
\bea \rho^2=\frac{u^2}{v},  \eea 
and the same is true for  $\tau_{11}= \tau_{11}(\rho).$
The function $\tau_{11}(\rho)$ has the following
singularities: 
 \beq \rho^2=4 \rightarrow v=\frac{1}{4} u^2; \qquad  \rho^2=-4 \rightarrow v=-\frac{1}{4} u^2; \eeq 
\beq \rho=\infty \rightarrow v=0. \eeq Thus around the
points $+2,-2,2i,-2i,\infty$ the function $\tau_{11}(\rho)$ has
non-trivial monodromies, which correspond to the $U_1(1)\subset
SU(2)$ charges  of the previous section.

\begin{figure}[ht]  
\begin{center}
\epsfig{file=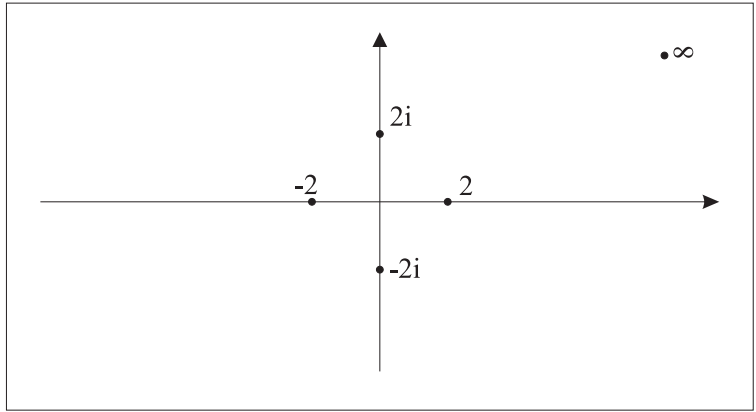,   width=10cm}
\end{center}
\caption{\footnotesize  The $\rho$ plane.} 
\end{figure}

These nontrivial  monodromies in the $\rho$ plane must be related to the massless states found earlier. 
We extract therefore   the magnetic and electric charges related to 
$U_1(1) \subset SU(2) $   from Table \ref{monopcharges}    and determine  the corresponding $2 $ by $2$ monodromy matrices, by using the
formula Eq.(\ref{charges}).    This gives the result shown in   Table \ref{ques}.  Note that these coincide with  the two by two  submatrices,
referring to the first and third rows and columns,    taken  out of the complete monodromy matrices found earlier (and listed in Appendix A)
\footnote{Although the matrices in Appendix A refer to the basis before the transformation Eq.(\ref{newbas}),  these
submatrices remain invariant:  ${\tilde m_1} =m_1$; ${\tilde q}_1= q_1$,   since  $q_2=0$. }.

\begin{table}[h]  
\begin{center}
\begin{tabular}{|l|l|l|} \hline Section & Charge & Matrix\\
\hline $M_1,M_4$ & $(\pm 1,0)^4$ & $
 \pmatrix{1& 0 \cr -4 & 1 }$
 \\ $A_2, A_5$ &$(\pm 1,\mp 1)^4$ & $\pmatrix{-3& 4 \cr -4 & 5 }$\\
$M_2,M_5$ &$(\pm 2,\mp 1)$& $\pmatrix{-1& 1 \cr -4 & 3 }$
\\ $A_3,A_6$ & $(\pm 2,\pm 1)$& $\pmatrix{3& 1 \cr -4 & -1 } $\\
$M_3,M_6$&$(0,\pm 1)$& $\pmatrix{1& 1 \cr 0 & 1 }$ \\
 $A_1,A_4$ & $(\pm 4,\mp 1)$ & $\pmatrix{-3& 1 \cr-16 & 5 }$ \\
\hline
\end{tabular}  
\end{center}
\caption{\footnotesize   Simplified monodromy matrices and massless charges} 
\label{ques}
\end{table}

\begin{figure}[h]
\begin{center}
\epsfig{file=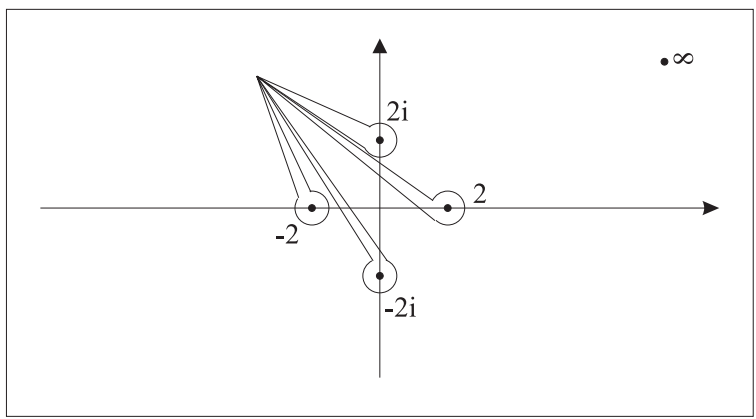,   width=10cm}
\end{center}
\caption{\footnotesize   Monodromies in the $\rho$ plane.}    
\label{monorho}
\end{figure}

To  assign  the various matrices of Table \ref{ques}  to the  monodromies around the  singularities in the $\rho$ plane, we have to take into
account   the consistency condition   (see Fig. \ref{monorho})  
\beq M(+2 i) M(+2) M(-2 i) M(-2)=M(\infty), \eeq
 where $M(+2 i)$ is the monodromy matrix around the point $+2 i$, etc.  
It turns out that the solutions is not unique.   By making use of the relations
\bea  &&   M_6 A_6  M_1=M_3 A_3 M_4= -{\bf 1}, \qquad   
A_1 M_2 M_1 =A_4 M_5 M_4=-{\bf 1}, \non \\  && 
M_3 M_2 A_2=M_6 M_5 A_5=-{\bf 1}   \eea  
which hold among the matrices of Table \ref{ques},  it is   possible to construct  different    solutions,    given in various columns of
Table \ref{bbbb}.

\begin{table}[h] 
\begin{center}
\begin{tabular}{|l|l|l|l|l|l|l|l|l|}
\hline $ M(+2 i)$ & $M_6 $
&$A_1$ &$M_3$ & $M_3 $
&$A_4$ &$M_6$&$M_6$ &
$\ldots$ \\
\hline $M(+ 2)$ & $A_6$
&$M_2$ &$M_2$ &$A_3$ &
$M_5$ &$M_5$ &$A_6$ & $\ldots $\\
\hline $M(- 2 i)$ & $M_3 $
&$A_4$ &$M_6$ & $M_6 $
&$A_1$ &$M_3$&$M_3$ &
$\ldots$ \\
\hline $M(- 2)$ &$A_3$
&$M_5$ &$M_5$ &$A_6$ &
$M_2$ &$M_2$ &$A_3$ & $\ldots $\\
\hline $M(\infty)^{-1}$ & $M_1 M_4   $
&$M_1M_4 $ &$A_2 A_5 $ &
$M_4 M_1$ &$M_4 M_1 $&$A_5 A_2 $ &$M_1 M_4 $ &
$\ldots $ \\
\hline
\end{tabular}
\end{center}
\caption{\footnotesize Different sections of the singularities describing the same physics}
\label{bbbb}    
\end{table}

\begin{figure}[h]
\begin{center}
\epsfig{file=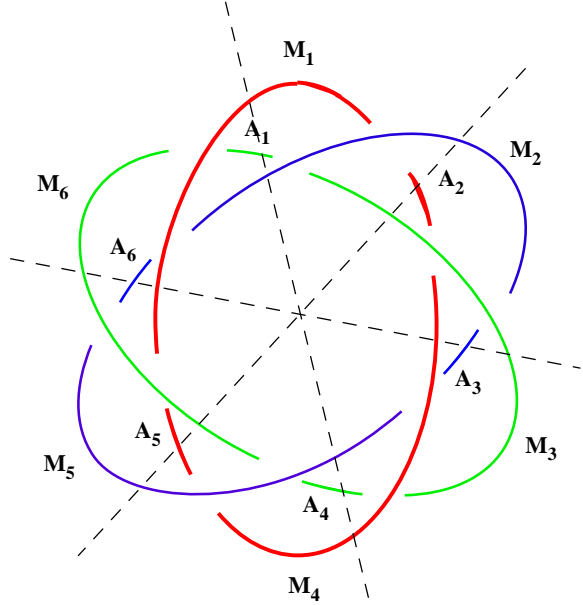,    width=8cm}
\end{center}
\caption{\footnotesize The linked rings cut in different sections. } \label{knotcut}
\end{figure}

A crucial observation is that  in the $(u,v)$  space there are infinite number of copies of  $\rho$ plane, 
corresponding to different phases of $\epsilon$. Namely,  
by  varying the phase of $\epsilon$    the
linked rings   in Fig.  \ref{knot}  are    cut   in different sections  (different copies of $\rho$ plane).   See Fig. \ref{knotcut}. 
It is therefore natural to identify  the set of singularities in each section  with  the  monodromy matrices of a given column   in Table
\ref{bbbb}.     Note that at constant
$\epsilon$, 
$\rho
\rightarrow -\rho $ corresponds to $\{v \rightarrow v, u 
\rightarrow -u \}$   (reflection with respect to the origin in Fig.\ref{knotcut} ):  this explains the 
appearance of a pair of particles  in each section with opposite charges with respect to  $U_1(1) $.     They belong  to a
doublet of $SU(2)$.

\begin{table}[h]
\begin{center}
\begin{tabular}{|l|l|l|l|l|l|l|l|}
 \hline $+2 i$  &$(0,1) $ &$(4,-1)$
&$(0,1)$ &
$(0,-1) $ &$(-4,1)$ & $(0,-1)$ &  $  \ldots $    \\
 \hline $ 2$ &$(2,1)$ &$(2,-1) $
&$(2,-1)$ &$(-2,-1)$ &
$(-2, 1)$ &$(-2, 1)$ & $\ldots $  \\
\hline $-2 i$  &$(0,-1) $ &$(-4, 1)$
&$(0,-1)$ &
$(0,1) $ &$(4,-1)$ &  $(0,  1) $ &   $  \ldots $  \\
\hline $-2$ &$(-2,-1)$ &$(-2, 1)$
&$(-2, 1)$ &$(2,1)$ &
$(2,-1)$ & $(2,  - 1)$ &  $\ldots $  \\
\hline $\infty$ &$(\pm 1,0)^4$ &$(\mp 1,0)^4$
&$(\pm 1,\mp 1)^4$ &
$(\mp 1,0)^4$ &$(\pm 1,0)^4$& $(\mp 1, \pm 1)^4$  
&  $\ldots $ 
\\
\hline
\end{tabular}
\end{center}
\caption{\footnotesize  The same as Table \ref{bbbb} but with charges. }
\label{aaaaa}     
\end{table}

The main point of this  discussion  is that the three different columns of charges (note the three-column  periodicity   in Table \ref{bbbb}
and Table
\ref{aaaaa})  
can be interpreted as   three different descriptions of the same physics, differing only by the redefinition of the charges.  
 Indeed, one  can pass from the first column ¡to the 
second by a $SL(2,Z)$ transformation    $p_1=\pmatrix{-1& 4 \cr 0 & -1 } $;
from the second to the third by   $p_2=\pmatrix{-1& -4 \cr 1 & 3 }
$; from the third to the first by   $p_3=\pmatrix{1& 0 \cr 1 &
1}$.

And this suggests  how to solve  the problem of non-uniqueness of the 
 $u, v \to 0$  limit, hence of the superconformal theory at hand. 
 We define,    as was done  in \cite{AD},        the superconformal limit by  
\beq    \epsilon \to 0,  \qquad     \rho \to 0,  
\eeq
i.e.,  $  \rho \to 0$  first. 
 Note that $\rho=0$  is the unique  point where  the symmetry of the 
equation Eq.(\ref{simm})   for the singularities, $\rho \to - \rho$,    is maintained.

 At $\rho=0$,  we find  that the four
colliding branch points are  at  $x_1= 2 i,\,\,x_2=-2 i,\,\,x_3=2,\,\,x_4=-2$ (in the unit of $\epsilon$): they   form a square.
 In order to apply  
(\ref{relation}),  we must appropriately identify  these four points,  consistently with the charges given  in the first column of Table 
\ref{aaaaa}.          

We note that  as  $\rho$   moves to  $2 i $ the branch points $x_3$ and $x_4$ coalesce.  As the monodromy matrix around $\rho=2  i $
implies the  massless particle there to have  the charge $(0,1)$,    we must assign  the   $\beta_1$  cycle to  encircle    the 
pair of branch points   $x_3$ and $x_4. $   An analogous consideration about the monodromy around $\rho= +\infty$, where the branch points
$x_1$ and $x_4$ coalesce,  giving rise to massless monopoles,  suggests that the $\alpha_1$ cycle defining the magnetic charge  encircle 
$x_1$ and $x_4$   (see Table  \ref{aaaaa}).  
We identify then  
\beq   e_3=   { 1\o   x_4 - x_2}  \to b , \qquad e_2=    { 1\o   x_1 - x_2} \to  c, \qquad e_1=   { 1\o   x_3 - x_2} \to  d, 
\eeq
in Eq.(\ref{relation}).   
We obtain at $\rho \to 0$  \bea { 1\o 2}  =\frac{ \theta_{00}^4
(0,\tau_{11})}{\theta_{10}^4(0,\tau_{11})},\label{modular}\eea
 which has the following solutions:
 \beq \tau_{11}=\frac{\pm 1+i}{2}, \qquad  \frac{\pm 3+i}{10},   \qquad   \ldots \label{tau}  \eeq
Other solutions of Eq.(\ref{modular}) can be found by acting repeatedly 
\beq   \tau \to  \tau+2; \qquad 
  \tau \to  {\tau  \o   1-2 \tau  } \eeq
on the solutions (\ref{tau}).

\section{ Low-Energy Physics at  a Renormalization-Group  Fixed Point \label{sec:LEP}}

The low-energy theory at the sextet vacua has a natural interpretation  as  a superconformal  $SU(2)\times U(1)$ gauge
theory,    with 4 magnetic monopole  doublets  $({\tilde m}_1,{\tilde q}_1) =  (\pm1, 0)$, a non-abelian electric doublet with charges
$({\tilde m}_1,{\tilde q}_1)=(0,\pm1)$   and
 a non-abelian dyon doublet with charges
$({\tilde m}_1,{\tilde q}_1)=(\pm2,\pm1)$. 
See the first  column of Table \ref{aaaaa}.  
The four magnetic doublets  have the second  ($U_2(1)$)   abelian
magnetic charge equal to ${\tilde m}_2=1$; the other two doublets   have abelian
magnetic charge equal to   ${\tilde m}_2=2$. 

The $SU(2)$ factor defines an interacting conformal theory. The
$\beta$ function cancellation occurs in the following manner.
The four magnetic monopole doublets  cancel   the contribution of the dual $SU(2)$  gauge
bosons and of their supersymmetric partners as in a local ${\cal N}=2,$    $SU(2)$ gauge theory with $n_f=4$. The non-trivial part of
the cancellation occurs  between the non-abelian electric  and    dyonic doublets.   By considering the 
$U(1)$ subgroup of the $SU(2)$,     their contribution to the first term of the  beta function cancel if \cite{AD}: \bea \sum_i (q_i+m_i
\tau)^2 =0.\label{cancellation} \eea
With the dyon charges $({\tilde m}_1,{\tilde q}_1)=(0,\pm1), (\pm2,\pm1)$  at our disposal, this works  if 
\beq
1 + (2\, \tau  +1)^2=0, \qquad    .^.. \quad     \tau^*=\frac{-1+i}{2}.  \eeq 
Note that this value of   critical coupling constant  is precisely (one of)  the value(s)   one finds from the behavior of  the small torus, 
(\ref{tau}). 

As we  change the basis  of our  theory by moving to  another section, e.g.,  corresponding to the second column of Table \ref{aaaaa},  
we redefine  the homology cycles defining   $a_{Di}, a_{i}$'s.   The coupling constant $\tau$ is accordingly  transformed:
\beq \tau^*   \rightarrow \frac{-\tau^*}{-4 \tau^*-1} =   { 3 + i \o 10},   \eeq  
which is again the one following from the limiting behavior of the curve, (\ref{tau}).   The condition for the beta function cancellation
Eq.(\ref{cancellation})   with charges  $({\tilde m}_1,{\tilde q}_1)=(\pm 4,\mp 1), (\pm 2,\mp 1), $  precisely leads to 
$\tau^*=    { 3 + i \o 10}$,    meaning that   the infrared fixed-point condition is satisfied independently of  the basis chosen to describing the
theory. 

Analogouly,  the transformation $\tau \rightarrow \frac{3 \tau-1}{4 \tau-1 }$ allows to  go from the second to the third basis, and 
 $\tau \rightarrow \tau-1$  to go from the third  back to the first.    The charges of massless fields and the critical coupling constant 
are transformed, but  the condition for the infrared fixed point  is  always satisfied. 

\begin{table}
\begin{center}
\begin{tabular}{|l|l|}
\hline First Basis & $\tau^*=\frac{-1+i}{2}$ \\
\hline Second Basis & $\tau^*=\frac{3+i}{10}$ \\ 
\hline Third Basis & $\tau^*=\frac{1+i}{2}$ \\
\hline
\end{tabular}
\end{center}
\caption{\footnotesize Critical coupling constant}    
\end{table}

We find  thus  a natural but a highly non-trivial     generalization of  the Argyres-Douglas 
infrared fixed-point theory  to one with a non-Abelian gauge symmetry.

\section { Superconformal Vacuum as Limit of Six Colliding Vacua}

The superconformal   limit may be approached by first breaking it explicitly by unequal bare quark masses,
by identifying the six nearby singularities $U_k, V_k$  ($k=1,2,\ldots 6$)   where the curve has the  form, 
\beq     y^2 =     (x^3  -   U  x  -
V )^2  -  \prod_{ a=1}^4    (x+m_a ) =  (x- \alpha)^2 (x- \beta)^2  (x-\gamma) (x- \delta), 
 \eeq
and then by considering  the limit of equal mass. In this limit, 
these  singularities   coalesce  and  becomes the conformal
vacuum.  

Each of the theories before the equal mass limit is taken   is a local   $U(1)^2$  gauge theory, with precisely two massless
hypermultiplets, each of which carrying  only one of the $U(1)$  charges.       There  are  in all    twelve massless hypermultiplets, and it is
 tempting to identify these degrees of freedom with the twelve massless particles found in the $m_i=0$   theory ($4+1+1$ 
doublets of the effective $SU(2)$ gauge group).

\begin{figure}[h]
\begin{center}
\epsfig{file=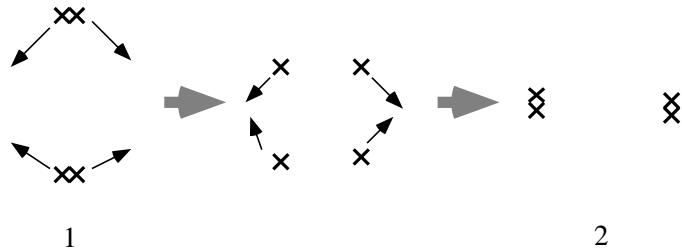,      width= 9  cm}  
\end{center}
\caption{\footnotesize A schematic representation of a nontrivial 
rearrangement of the four nearby branch points in the $x$ space, as  one moves from a singularity 
$(U_1,V_1)$ to another  $(U_2,  V_2).$  } \label{Rearrange}
\end{figure}

\begin{figure}[h]
\begin{center}
\epsfig{file=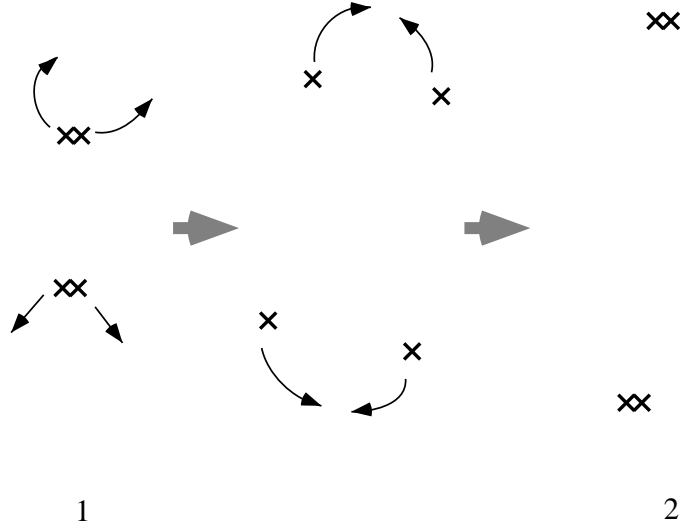,      width= 9  cm}  
\end{center}
\caption{\footnotesize A trivial 
rearrangement of the  branch points, as  one moves from a singularity 
 to another, having relatively local massless particles. } \label{trivial}  
\end{figure}

A partial support  comes  from the observation  that the massless states at one singularity and those at another singularity can be 
relatively non-local to each other.  That this does occur can be explicitly verified by studying the movements of the four 
branch points near the origin ($|m_i| \ll \Lambda$),  of the curve
\beq  y^2 =  \prod_{i=1}^{3}  (x-\phi_i)^2 -  \prod_{ a=1}^4    (x+m_a ) \equiv    (x^3  -   U  x  -
V )^2  -  \prod_{ a=1}^4    (x+m_a ),
 \eeq
as one moves from a singularity  $(U_1,V_1)$ to another  $(U_2,  V_2), $   by a numerical method.  An example of such a non-trivial rearrangement of the
branch points is illustrated in Fig. 
\ref{Rearrange}.   There are also pairs of singularities which correspond to relatively local massless states (Fig. \ref{trivial}). 
This is consistent with the charges present in  the theory  (see the first column  of Table \ref{aaaaa}).

 \vskip 0.5cm

Nonetheless,   there are reasons to believe that the mechanism of 
confinement in the superconformal theory, deformed by the adjoint mass term $\mu \, \Tr \, \Phi^2$,  cannot be understood this way. 
In fact, consider adding the  term $\mu \, \Tr \, \Phi^2$  in each of the six   vacua   $U_i, V_i$   with  a $U(1)^2$   gauge  symmetry.  The superpotential
of the theory has the form,
\beq  {\cal P} =   \sum_{i=1}^2   \sqrt {2}   \, A_{D_i}  \,   M_i  {\tilde M}_i  +  \mu \,  U (A_{D1}, A_{D2})  + {\hbox {\rm mass terms }} 
\label{superp}  \eeq
 The standard argument shows  that the $k$-th  vacuum is characterized by the equations
\beq     \mu = -  \sum_{i=1}^2  {\de a_{D i } \o \de U}  \, M_i  {\tilde M}_i   ; \qquad   0    =  \sum_{i=1}^2  {\de a_{D i } \o \de V } \,   M_i  {\tilde M}_i ,
\eeq
where   $ {\de a_{D i } \o \de U} $ and   $ {\de a_{D i } \o \de V} $  are given by the period integrals of the holomorphic differentials,
see Appendix C.   Other equations following from   (\ref{superp})  tell that    the  $M_i, \,  {\tilde M}_i $  fields are massless.    It
can  be seen easily that  in the equal mass limit (superconformal limit), in which the  four branch points come together, 
\beq      {\de a_{D1  } \o \de U}  \to \infty, \qquad     {\de a_{D2  } \o \de U}  \to \infty, 
\label{infinite}\eeq
hence
\beq    \bra M_1 \ket,  \,\,  \bra {\tilde M}_1 \ket,     \,\,  \bra M_2 \ket,  \,\,  \bra {\tilde M}_2  \ket  \,\,   \to  0.  
\label{tend}\eeq
  The same holds for all massless hypermulltiplets  at the six  ${\cal N}=1$ vacua.   This is analogous to  the phenomenon
discussed in \cite{GVY}  at the Argyres-Douglas point of  ${\cal N}=2,$    $SU(2)$  gauge theory with $n_f=1$.

 \section{ Non-Abelian vs Abelian Confinement Mechanism: QCD }

This brings us to an apparently  paradoxical situation.    The consideration of the equal mass limit starting from the unequal mass cases 
allow us to find the candidate degrees of freedom and identify them with those  which appear in the massless
 theory studied in the earlier sections.  This also seems to allow  to study  the effect of  the ${\cal N}=1$ perturbation and the ensuing 
confinement and  dynamical symmetry breaking, in terms of   the local effective actions valid in each of the six vacua.  
However, in the equal-mass limit,  all  condensates are found to  vanish   (at  finite $\mu$).  

On the other hand,  a detailed study of the  ${\cal N}=2$, $SU(3)$  gauge theory with $n_f=4$ at  a  large adjoint mass $\mu$,     
made in \cite{CKMP}, shows that  at the sextet vacua confinement and dynamical symmetry breaking
\beq    SU(4) \times U(1) \to    U(2) \times U(2)     \label{sbp} 
\eeq
do  take place.  By supersymmetry and holomorphy (physics depends on $\mu$,  not on $|\mu|$),  we {\it know }   that the same result holds
at small
$\mu$. 
 But then which is the order parameter
of the confinement/dynamical symmetry breaking? 
  
The fact that  these (almost) superconformal  theories are   non-Abelian gauge  theories with finite or infinite
  couplings, seems to show the way out. 
In the preceding discussion Eq.(\ref{superp}) - Eq.(\ref{tend}), the effects of the strong $SU(2)$ interactions, which is probably  dominant, 
 are entirely  neglected.     It is perfectly possible that  the four magnetic monopole doublets of   $SU(2)$   discussed in Sec \ref{sec:LEP},  ${\cal
M}_{\alpha}^i,$   ($\alpha=1,2$,   $\, i=1,\ldots, 4$),    condense due to the $SU(2)$ interactions  upon  $\mu$ perturbation,
\beq     \bra    {\cal M}_{\alpha}^i  \, {\cal M}_{\beta }^j \ket  =     \epsilon_{\alpha   \beta}  \,   C^{ij} \ne 0, 
\label{condensate}\eeq  
 where   $C^{ij}$ is  antisymmetric in the flavor indices.  Such a structure is consistent with  the known symmetry breaking pattern,
Eq.(\ref{sbp}).\footnote{One of the $U(1)$ is a combination of the gauge $U(1)$ group and a subgroup of the flavor $SU(4)$.  
 That Eq.(\ref{condensate})  leaves
the other   $U(1)$ global symmetry    is not so obvious.  There is however  a natural mechanism for the  quantum quenching of the 
baryonic   $U(1)$
  charge for magnetic monopoles
\cite{KT,CKMP}.   }

   We are thus led naturally to the conclusion that   the microscopic mechanism of confinement and dynamical symmetry 
breaking  in the   almost superconformal vacua  such as the one under consideration,    
should be  essentially different from the Abelian confinement mechanism  found in the original Seiberg-Witten work \cite{SW1,SW2},
and involves strongly interacting  non-Abelian magnetic degrees of freedom.\footnote{Analogously,  we believe  that 
 the vanishing of the monopole  and charge condensates at the Argyres-Douglas point  found in  Ref.~\cite{GVY}  means the inadequacy of 
these degrees
of freedom  as the order parameters, rather than signalling the deconfinement.  In fact,  in that
model (an ${\cal N}=2,$  $SU(2)$ theory with $n_f=1$),  the standard squark condensate $\bra Q {\tilde Q} \ket$    remains finite at the
Argyres-Douglas point,  which cannot be simply expressed in terms of the monopole/dyon degrees of freedom.}

The final  aim of our efforts is to    understand  the microscopic mechanism of confinement in    the  standard QCD. 
As there are other reasons to believe that  the 
   model of confinement by weakly-coupled  magnetic particles  is not   a good model    for  QCD \cite{AK},  
it is exceedingly  interesting    that   in the class of confining vacua   based on deformed superconformal theory   we have a  new 
mechanism  of confinement, involving    strongly interacting magnetic degrees of freedom.  It has been  noted also    \cite{CKMP}    that
the pattern of dynamical symmetry breaking in these SCFT based  vacua  was  most reminiscent of  what happens   in QCD.

Another  line of thought  along   a recent work \cite{BK},   
appears to lead to a similar conclusion.  
  There, the appearance  of the non-Abelian monopoles as weakly coupled 
low-energy degrees of freedom  in the  $r$-vacua of ${\cal N}=2$  supersymmetric QCD,  $r < { n_f \o 2}$,   was understood as a consequence
of  the flipping of  the sign of the  beta function coefficient:     
\beq     b_0^{(dual)} \propto   -  2 \, r  +   n_f  >  0,   
\label{betadual}   \eeq
for dual $SU(r)$ theory,   while in the fundamental  $SU(n_c)$   theory
\beq     b_{0} \propto  -  2 \, n_c +    n_f  <     0.     
\label{betafund}   \eeq
The dressing of   the non-Abelian monopoles with the flavor quantum number is a neccessary condition for this to happen.   Indeed,  in a pure 
 ${\cal N}=2$ Yang-Mills theory, or on a generic point of the moduli space in ${\cal N}=2$  supersymmetric QCD,  where this is not possible,   only 
monopoles of Abelian variety  appear as the   low-energy degrees of freedom.    

The appearance of Abelian monopoles as  long-distance    physical  degrees of freedom   is in itself consistent as $U(1)$ theories
are  infrared free.  However,  this phenomenon  seems to be always accompanied by dynamical gauge symmetry breaking ({\it dynamical
Abelianization}),   with  characteristic  signals of enriched  spectrum of Regge trajectories, etc.

In the standard QCD, the coefficient $2$  in front of the color multiplicity $n_c=3$ or    $r (\le 3)$    in  Eq.(\ref{betafund}), Eq.(\ref{betadual}), is
replaced by
$11$.  It is clearly very difficult  to realize, having at our disposal only  a few light flavors,   the required sign flip.  It is even more so, since the
monopoles, being scalars,  contribute less than the quarks in the original Lagrangian.   On the other hand, there are no phenomenological
indications that dynamical Abelianization takes place in the real world of strong interactions. 

We are naturally  led to a picture of confinement in QCD, still largely to be clarified,   involving strongly-interacting non-Abelian magnetic
degrees of freedom.

\section*{Acknowledgments}

The authors acknowledge  useful  discussions  with many friends and collegues  which contributed   to the progress of this work,  
especially with   Stefano Bolognesi,  T. Eguchi, Jarah Evslin,   E. Guadagnini,  P.S. Kumar,  W. Lerche, M. Mintchev,
   A. Ritz,    D. Tong,    A. Vainshtein  and     A. Yung.  Some of the algebraic
analysis needed in this work  have been carried out  by using   Mathematica 4.0.1  (Wolfram Research).

\appendix

\section{Monodromy Matrices  }

The set of  the twelve    matrices:
\bea
&&   M_{1}  =  \pmatrix   { 1 & 0& 0  & 0 \cr
0& 1 & 0& 0 \cr   -4   & 0& 1& 0\cr
 0& 0& 0&  1  } , \qquad  M_{2}  = \pmatrix   { -1 & 0& 1  & 0 \cr    0& 1 & 0& 0 \cr   -4 & 0& 3& 0\cr
 0& 0& 0&  1  }, \non \\
&&   M_{3}  =  \pmatrix   { 1 & -1 & 1  & 0 \cr    0& 1 & 0& 0 \cr   0  & 0&1 & 0\cr
 0 & -1  & 1 &  1  }, \qquad  M_{4}     \pmatrix   {1 & 0 & 0 & 0 \cr    0& 1 & 0& 0
\cr  -4 & 4  & 1& 0\cr 
4 & -4&  0  &  1  },    \non \\
&&   M_{5}  =  \pmatrix   { -1 & 2 & 1  & 0 \cr    0& 1 & 0& 0 \cr   -4 & 4 & 3  & 0\cr
 4& -4 & -2  &  1  }, \qquad  M_{6}  = \pmatrix   {1 & 1& 1  & 0 \cr    0& 1 & 0& 0
\cr   0 & 0& 1& 0\cr
 0& -1& -1 &  1  }, 
  \non \\
&&   A_{1}  =  \pmatrix   { -3 &  1 & 1  & 0 \cr    0& 1 & 0& 0 \cr  -16  & 4& 5& 0\cr
 4 & -1 & -1  &  1  }, \qquad  A_{2}  =  
 \pmatrix   {-3 & 0 & 4 & 0 \cr    0& 1 & 0& 0
\cr -4 & 0& 5 & 0\cr 
 0& 0& 0   &  1  },  \non \\
&&   A_{3}  =  \pmatrix   { 3 & -2 & 1  & 0 \cr    0& 1 & 0& 0 \cr   -4 &4& -1& 0\cr
 4& -4& 2 & 1  }, \qquad  A_{4}  =  \pmatrix   {-3 & 3& 1  & 0 \cr    0& 1 & 0& 0
\cr   -16 &  12 & 5 & 0\cr 
 12& -9 & -3 &  1  },  \non \\
&&   A_{5}  =
 \pmatrix  {-3 & 4 & 4  & 0 \cr    0& 1 & 0& 0 \cr   -4 & 4& 5 & 0\cr
 4& -4& -4 &  1  }, \qquad  A_{6}  =  \pmatrix   {3 & 0 & 1  & 0 \cr    0& 1 & 0& 0
\cr   -4  & 0& -1 & 0\cr 
 0& 0&0  &  1  },  \non \\
\eea
satisfy all the conjugation  relations, Eqs.(\ref{conju}).  
Note also that  
\beq   M_{1}  =({\tilde M}_1)^4,\quad  M_{4}  =({\tilde M}_{4})^4, \quad 
A_{2}  = ({\tilde A}_2)^4, \quad  A_{5}  = ({\tilde A}_5)^4, \eeq
with 
\bea   &&   {\tilde M}_1=   \pmatrix   { 1 & 0& 0  & 0 \cr
0& 1 & 0& 0 \cr   -1 & 0& 1& 0\cr
 0& 0& 0&  1  },  \quad  {\tilde M}_{4}=   \pmatrix   {1 & 0 & 0 & 0 \cr    0& 1 & 0& 0
\cr  -1 &  1  & 1& 0\cr 
1 & -1&  0  &  1  },     \non \\  &&     
 {\tilde A}_2=  
 \pmatrix   {0 & 0 & 1 & 0 \cr    0& 1 & 0& 0
\cr -1 & 0&2& 0\cr 
 0& 0& 0   &  1  },  
 \quad  {\tilde A}_5 =  
 \pmatrix  { 0 & 1 & 1  & 0 \cr    0& 1 & 0& 0 \cr   -1 & 1& 2 & 0\cr
 1& -1& -1 &  1  }. 
\eea
 The use of the formula Eq.(\ref{charges})
then yields the charges Eq.(\ref{chargesbis}).

\section  {Inversion formula  }

A genus one curve can be parametrized in terms of a Weierstrass function  as 
\beq y =  {d {\cal P}(z;
\omega_1, \omega_2)    \o dz },   \qquad   x=  {\cal P} (z;
\omega_1, \omega_2), \eeq
which satisfies 
\beq   y^2= 4 \,x^3 - g_2 \, x - g_3 =   4 \, (x-e_1) \, (x-e_2) \, (x-e_3),
\eeq
where
\beq  e_1=    {\cal P}({\omega_1 \o  2}), \quad e_2=    {\cal P}({\omega_2 \o 2}), \quad e_3=    {\cal P}({\omega_1 + \omega_2 \o 2}).   
\eeq     
The  shape parameter  is given by 
$\tau= {\omega_1 \o  \omega_2},$
\beq \omega_1 =  \oint_{\alpha} { dx \o y}, \qquad   \omega_2 =  \oint_{\beta} { dx \o y},   
\eeq
where $\alpha$ and $\beta$ cycles  encircle the pairs of branch points  $(e_3, e_2)$ and    $(e_3, e_1),$
respectively.   With these definitions, 
the inversion formula is: 
\beq   \lambda^{-1}(\tau)  =  { \theta_3^4(0, e^{i \pi \tau}) \o   \theta_2^4(0, e^{i \pi \tau})}   =   {e_3 - e_2 \o e_3 - e_1}. 
\eeq

\section  {Period Integrals  at the Sextet Singularities  }

According to the by  now standard result \cite{curves}, the derivatives of $A_{D i}$ and $A_{i}$'s  with respect to 
the vacuum parameters  $U=  \bra \Tr \, \Phi^2 \ket$ and   $V=  \bra \Tr \,\Phi^3 \ket  $  are to be identified with the period 
matrices,
\bea   {d A_{D1} \o  dU } &=&  \oint_{\alpha_1}  { dx \o y},   \qquad    {d A_{D2} \o  dU } =  \oint_{\alpha_2}  { dx \o y},   \non \\
{d A_{D1} \o  dV } &=&  \oint_{\alpha_1}  {  x \,dx \o y},   \qquad    {d A_{D2} \o  dV} =  \oint_{\alpha_2}  {x  \,dx \o y},   \non \\
{d A_{1} \o  dU } &=&  \oint_{\beta_1}  { dx \o y},   \qquad    {d A_{2} \o  dV } =  \oint_{\beta_2}  { dx \o y},   \non \\
{d A_{1} \o  dV} &=&  \oint_{\beta_1}  {x\,dx \o y},   \qquad    {d A_{2} \o  dV } =  \oint_{\beta_2}  { x\, dx \o y},   \non \\
\eea
where
\beq   y^2  =    (x^3  -   U  x  -
V )^2  -  \prod_{ a=1}^4    (x+m_a )  = \prod_{i=1}^6   (x- e_i).  
\eeq
At the superconformal vacua of our interest,   four of the branch points $e_i$, $i=1,2,3, 4$  coalesce whereas
two others are at $ e_5, e_6 = \pm 1$.   By our choice, $\alpha_1$ cycle encircles the branch points  $e_1$  and  $e_4$;  $\beta_1$ cycles encircle the
branch points  $e_3$  and  $e_4$;  $\alpha_2$ cycle encircles all the branch points  $e_1,\,e_2,\,e_3,\, e_4,$ and   the large $\beta_2$ cycle
encircles the points $e_2$  and  $e_5.$  Near the   SCFT point, 
\beq    {d A_{D1} \o  dU } \simeq   { 1\o \sqrt {e_5 \, e_6}}  \, 2 \int_{e_1}^{e_4}   \, { dx \o \sqrt   {(x-e_1)(x-e_4)(x-e_2)(x-e_3)} }
\stackrel{SCFT}{\longrightarrow }\infty, 
\eeq
etc.    It is easy to verify Eq.(\ref{infinite}) similarly. 

\end{document}